\documentclass[aps,prd,floatfix,nofootinbib,amssymb,preprintnumbers,showpacs,twocolumn]{revtex4}
\usepackage{graphicx}
\begin{document}
\newcommand{\tr}{\mbox{Tr}\,}
\title
{The $D_{sJ}^+(2317)$: what can the Lattice say?}
\author{Gunnar S.\ Bali}
\email{g.bali@physics.gla.ac.uk}
\affiliation{Department of Physics \& Astronomy, The University of Glasgow, Glasgow G12 8QQ, Scotland}
\date{\today}
\begin{abstract}
We present lattice results on the scalar $D_s$ meson and comment
on the $D_{sJ}^+(2317)$ state recently discovered by BaBar
and confirmed by CLEO, in view of a series of theoretical claims
and counter claims. Lattice predictions in the static limit
indicate larger masses
than observed for a scalar quark model state.
Finite $c$ quark mass corrections seem to further enlarge this discrepancy,
in support of a non quark-antiquark-state interpretation of experiment.
\end{abstract}
\maketitle
\section{Introduction}
Recently the BaBar Collaboration announced the discovery of a
$D_s^+$ positive parity meson at $2317\pm 3$~MeV and a width smaller
than 10~MeV in the $D_s^+(1969)\pi^0$ channel~\cite{Aubert:2003fg}.
No spin assignment has been made as yet but in view of the low mass
$J=0$ appears likely. For simplicity we shall refer to it as
the scalar or the $0^+$ $D_s$ meson.
This state was subsequently confirmed by the CLEO
Collaboration~\cite{Besson:2003jp}.
CLEO also reports the observation of another state near
2460 MeV in the $D_s^{*+}(2112)\pi^0$ channel
which is consistent with having $J^P=1^+$.
These discoveries triggered a series
of articles with different claims. In this note
we discuss the scalar state in view of recent lattice results, after
briefly summarising the different interpretations.

\section{Quark model or not?}
If one treats the charm quark as a heavy spectator, the spin and
angular momentum of the light antiquark can either couple to $j=\frac{1}{2}^-$
($l=0$) or to $j=\frac{3}{2}^+$ and $j=\frac{1}{2}^+$ ($l=1$). The interaction with
the spectator spin will then result in a pseudoscalar-vector
mass splitting for $j=\frac{1}{2}^-$, in $0^+$ and $1^+$ states
for $j=\frac{1}{2}^+$ and in $1^{\prime +}$ and $2^+$ states for $j=\frac{3}{2}^+$.
The two $1^+$ states can undergo mixing. The pseudoscalar
and vector $D_s$ states have been identified as
$D_s^+(1969)$ and $D_s^{*+}(2112)$, respectively. Then there is
a $D_{s1}^+(2536)$ state and a $D_{sJ}^+(2573)$ which, with the
likely spin assignment $J=2$ in the latter case, form the $j=\frac{3}{2}$ doublet.
The $j=\frac{1}{2}^+$ states can strongly decay into $DK$ and $D^*K$
and are expected to be
broad resonances. The new $D_{sJ}^+(2317)$ and the state
at 2.46~GeV might constitute the missing doublet, where at least
the former state,
which lies almost 40 MeV below the $DK$ threshold, is narrow.
Cahn and Jackson~\cite{Cahn:2003cw} interpret experiment in this way, in
the context of a potential model.

Barnes and collaborators~\cite{Barnes:2003dj} in contrast argue that this state
is most likely a $DK$ molecule since its mass is 160~MeV lighter
than other
potential model predictions which result in a mass around
2.48 GeV~\cite{Godfrey:wj} for the scalar  $P$-wave $D_s$.
Their argument is supported by the proximity to the $DK$ threshold
and they interpret this system as a generalisation of an $a_0/f_0(980)$
$K\overline{K}$ molecule. A four-quark interpretation
is also shared by Cheng and Hou~\cite{Cheng:2003kg}.
Szczepaniak~\cite{Szczepaniak:2003vy} argues in favour of a strong
$D_s\pi$ atomic contribution.

Van Beveren and Rupp~\cite{vanBeveren:2003kd} also liken this state with
the $a_0/f_0(980)$ but interpret it as a quark model state.
In their view the $a_0/f_0$ states are part of a low lying
scalar quark-antiquark nonet, together with a $\sigma(600)$ and
a $\kappa(800)$. Consequently, they postulate additional
scalar $D$ mesons. According to them, in both the $a_0/f_0$ and
the new $D_{sJ}^+$ systems, due to mixing with the $\overline{K}K$ or
the $DK$ continuum, respectively, the lowest scalar nonet is artificially
lowered with respect to the quark model expectation.

Bardeen et al.~\cite{Bardeen:2003kt} discuss the heavy quark limit.
They then follow Refs.~\cite{nowak,Bardeen:1993ae} and interpret
the $0^-$ -- $0^+$ splitting
in terms of chiral symmetry. The symmetry breaking scale
corresponds in leading order to the constituent quark
mass in the chiral limit~\cite{nowak} and has been estimated
to be~\cite{Bardeen:1993ae}
$\Delta M\approx 338$~MeV, a value that
is very close to the experimental splitting of $\approx 349$~MeV.
Chiral loops however will somewhat reduce the former expectation~\cite{nowak}.
Colangelo et al.~\cite{Colangelo:2003vg} share this picture and
Godfrey~\cite{Godfrey:2003kg} investigates the decays
that one would expect in the case of a quark-antiquark interpretation.

One should note that the vector-scalar
splitting, which vanishes in the heavy quark limit, is as large as 143~MeV
in the $D_s$ system, indeed an ${\mathcal O}(\Lambda/m_c)$ correction
to $\Delta M$. In view of this, we would not expect the static
approximation to be quantitatively correct for $D$ systems.
We also remark that the $0^+$ can be interpreted as a chiral partner
of the $0^-$, independent of the quark model content, as long as
isospin and strangeness agree. Unfortunately, most predicted decay rates
in many of the above pictures seem to be more dictated by the mass
and quantum numbers of the state than by its quark content. However,
in the case of an interpretation as a molecule or as part of an additional low
lying scalar nonet (or triplet), an extra quark-model scalar
state should still
exist above the $DK$ threshold. However, this might turn out to be
a rather broad resonance~\cite{Godfrey:wj}. In contrast, in a straight
$D_s$ interpretation there is no room for extra states other than
the $D_{s1}^{\prime +}(2460)$ and $D_{s1}^{+}(2536)$ between a
$D_{sJ}^+(2573)$ ($J=2$?)
and the newly discovered $D_{sJ}^+(2317)$ ($J=0$?). The chiral
heavy quark interpretation results in similar predictions
for $B$ systems~\cite{Bardeen:2003kt,nowak,Bardeen:1993ae},
which in principle can be checked experimentally.

\section{The static limit}
We will confront the new scalar state with lattice results in the static limit
in the quenched approximation and for $n_f=2$, before discussing finite
charm quark mass corrections.

In the static limit the $j=\frac{1}{2}$ and $j=\frac{3}{2}$ doublets will
be exactly mass degenerate.
We wish to calculate the $\frac{1}{2}^-$ and $\frac{1}{2}^+$ masses.
These can be extracted from the asymptotic large $t$ decay of the
two Euclidean correlation functions,
\begin{eqnarray}
C_{\pi}(t)&=&U_{0,t}\tr\left\{\frac{1+\gamma_4}{2}{M^{-1}_{0,t}}^{\dagger}\right\},\\
C_{\sigma}(t)&=&U_{0,t}\tr\left\{{M^{-1}_{0,t}}^{\dagger}\frac{1-\gamma_4}{2}\right\},
\end{eqnarray}
respectively.
We made use of the relation $M^{\dagger}=\gamma_5M\gamma_5$ for the
Wilson-Dirac operator $M$ and $\gamma_4\gamma_5=-\gamma_5\gamma_4$.
$U_{0,t}$ denotes the Wilson-Schwinger line, connecting the point
$({\mathbf x},t)$ with $({\mathbf x},0)$. The
spatial coordinate ${\mathbf x}$ is suppressed in $U$ as well as in $M$ and
the colour trace is implicit.

\begin{figure}[htb]
\includegraphics[width=8.4cm]{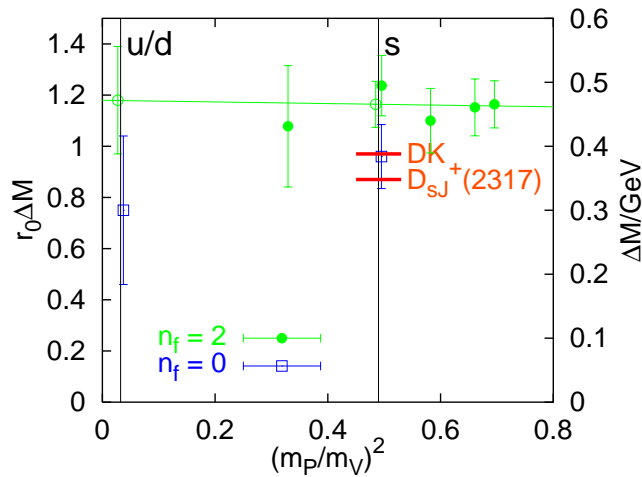}
\caption{
\label{figure} 
The $\frac{1}{2}^+$ -- $\frac{1}{2}^-$ splitting in the static limit, as a 
function of the light quark mass $\propto m_{P}^2/m_{V}^2$.
Open symbols denote inter- and extrapolations to physical up/down
and strange quark masses. Circles denote the $n_f=2$
case while squares denote the quenched approximation. The horizontal
lines are the experimental values for the $D_{sJ}^+(2317)$ --
$D_s^+(1969)$ splittings and the $DK$ threshold.}
\end{figure}

This static-light splitting has been
calculated in the quenched approximation by Michael and
Peisa~\cite{Michael:1998sg} with Wilson action at $\beta=5.7$
and $\beta=6.0$. The results are in agreement with earlier
references~\cite{Alexandrou:1992ti,Duncan:1994uq,Ewing:1995ih} at
additional
lattice spacings and
no significant
lattice spacing dependence has been observed: the fine structure splittings,
that strongly depend on short distance
physics~\cite{Bali:1998pi}, vanish by definition.
The quenched results,
extrapolated to up/down and strange quark masses are depicted in
Figure~\ref{figure} (open squares) where $r_0^{-1}\approx 400$~MeV.
The results are $\Delta M_s=384(50)$~MeV and $\Delta M_u=299(114)$ MeV,
respectively.
The splittings of the $\frac{1}{2}^-$ states with respect to the
$\frac{3}{2}^+$ states are 434(5)~MeV and 323(131) MeV for strange and up/down quarks,
roughly 50~MeV larger.
In contrast, the $D_{s1}^+(2536)$ is 120~MeV heavier than the scalar
$D_s$.

\begin{table}[hbt]
\caption{\label{table}
The static $\frac{1}{2}^+$ -- $\frac{1}{2}^-$ mass splittings
$\Delta M$
for $n_f=2$ sea quarks~\cite{Bali:2000vr} for different
hopping parameters $\kappa$ at $\beta=5.6$. The numbers in the last column are
subject to an additional 5~\% overall scale uncertainty.}
\begin{ruledtabular}
\begin{tabular}{ccccc}
$\kappa$&$r_0/a$&$m_P/m_V$&$\Delta M r_0$&$\Delta M/$MeV\\\hline
0.1560&5.11(3)&0.834(~3)&1.16 (9)&465(~35)\\
0.1565&5.28(5)&0.813(~9)&1.15(11)&460(~45)\\
0.1570&5.48(7)&0.763(~6)&1.10(13)&440(~50)\\
0.1575&5.89(3)&0.704(~5)&1.24(12)&495(~50)\\
0.1580&6.23(6)&0.574(13)&1.08(24)&430(110)
\end{tabular}
\end{ruledtabular}
\end{table}

Results with $n_f=2$ mass-degenerate flavours of Wilson sea quarks
have been obtained by the SESAM Collaboration~\cite{Bali:2000vr}
(circles) on slightly finer lattices. The data of Table~\ref{table}
do not exhibit any
visible light quark mass dependence. The interpolated values have been
obtained from a linear fit in $m_P^2$ (with tiny slope) and the
errors of the interpolation, that
have been conservatively
estimated by varying fit range and functional form, are dominated by
systematics.
The lattice spacing $a$ obtained from the phenomenological value
$a/r_0\approx a\times 400$~MeV is in agreement
with the
one obtained from $m_{\rho} a$, within errors~\cite{Bali:2000vr}.
The experimental lines in the figure correspond to masses,
relative to the pseudoscalar
state, a somewhat arbitrary choice since vector and pseudoscalar will
be degenerate in the static limit. The rationale behind this is
that Ref.~\cite{Bardeen:2003kt}
assumes the $1^+$ -- $0^+$ splitting to
be identical to the $1^-$ -- $0^-$ splitting.
The inclusion of sea quarks seems to result in the slightly increased value,
$\Delta M=468(43)(24)$~MeV for the $s$ quark system.
We do not expect a lattice spacing dependence of this number
in excess of the statistical
uncertainty, based on the quenched experience.

\section{Finite mass corrections}
Effects of the finite charm quark mass have only been investigated in the
quenched approximation. In particular three studies exist: two using
lattice NRQCD~\cite{Hein:2000qu,Lewis:2000sv}
to order $1/m^2$ and $1/m^3$, respectively (the leading
corrections are of order $\alpha_s/m$ in both cases), and
one using relativistic charm quarks~\cite{Boyle:1997rk}.
Both NRQCD results are consistent with each other.
The study of Hein {\em et al.}~\cite{Hein:2000qu}
has been performed at $\beta=5.7$ and $\beta=6.2$ for the $B_s$ and $B_d$
families and at $\beta=5.7$ for the $D_s$.
The relativistic study has been made at $\beta=6.0$ and $\beta=6.2$.
In the latter case we refrain from citing values for the $B$ meson
since the extrapolation of results obtained for heavy quark masses much
lighter than
the $b$ is not fully under control. In none of these cases statistically
significant lattice spacing effects have been observed and we display
the results for the $0^+$ -- $0^-$ splittings in Table~\ref{deltam}.

\begin{table}[hbt]
\caption{\label{deltam}
The $0^+$ -- $0^-$ mass splitting in the heavy-light system
for two sea quarks in the static limit
and in the quenched approximation, for the $B$ and $D$ systems in
NRQCD~\cite{Hein:2000qu} and for the $D$ system with relativistic
quarks~\cite{Boyle:1997rk}. The errors do not include uncertainties
in the overall scale which we estimate to be about 5~\% for
$n_f=2$. All numbers are in units of MeV.}
\begin{ruledtabular}
\begin{tabular}{c|c|cccc}
&$n_f=2$&\multicolumn{4}{c}{$n_f=0$}\\
&static&static&NRQCD&NRQCD&relativ.\\
&''&''&$h=b$&$h=c$&$h=c$\\\hline
$h\overline{s}$&468(43)&384 (50)&345(55)&465(50)&495(25)\\
$h\overline{d}$&472(85)&299(114)&370(50)& ---   &465(35)
\end{tabular}
\end{ruledtabular}
\end{table}

Note that while the NRQCD results for the $B$ systems agree with the respective
$n_f=0$ static limits, the splitting is enhanced in the $D$ system,
in agreement with the fully relativistic calculation. The relativistic
$D_s$ splitting is bigger by as much as $(29\pm 16)$~\% with respect
to the static limit. If we assume a similar increase for the case with
sea quarks we would expect a splitting of 600(110)~MeV for the
$D_s$ system, yielding the prediction
$m(D_{s0}^+)=2.57(11)$~GeV. The potential model of
Ref.~\cite{Godfrey:wj} predicts 2.48~GeV, while the
quenched results are 2.44(5)~GeV (NRQCD) and 2.47(3)~GeV (relativistic
$D$ quark), all significantly bigger than the candidate's mass
of 2.32~GeV.
The quenched lattice results for the $D_s$ system also indicate
a tiny $1^{\prime +}$ -- $1^+$ splitting, suggesting that the $1^+$ state should be
heavier than 2.46~GeV.

\section{Summary}
We calculate a scalar-pseudoscalar splitting of $\Delta M=468(43)(24)$~MeV
in the static limit for $n_f=2$ sea quarks, significantly larger than
the value 338~MeV suggested by a heavy quark constituent quark
model~\cite{Bardeen:1993ae} and larger than the quenched QCD value
$\Delta M=384(50)(20)$~MeV.
We also report a significant finite charm mass correction
that casts doubt onto na\"{\i}ve generalisations to the $B$ system.
Lattice predictions on the masses are
consistent with the quark model of Ref.~\cite{Godfrey:wj} and
incompatible with the new state observed by BaBar and CLEO.
We conclude that the $D_{sJ}^+(2317)$ might receive a large $DK$
component: the physics of this heavy scalar might indeed
resemble elements of that governing the $f_0/a_0(980)$ system.
If this is the case then the masses of the up and down quarks will play a major
r\^ole and simulations with non-mass-degenerate sea quarks are required.

Unfortunately, on the lattice the possibility of four-quark states
has so far only been addressed in the static limit where attraction
was reported in some channels~\cite{Michael:1999nq}.
In view of the new experimental candidate quenched simulations
of relativistic four quark molecules are urgent. To understand the
exact nature of the new state not only the spectrum but also predictions
of decay rates are required.
While lattice
calculations of strong decays are unfeasible, a study of
electro-magnetic decay rates is a possibility.\\

\noindent {\bf Notes added in proof} \\[.2cm]
The discovery of the two $D_s$ mesons has also been confirmed
by the Belle Collaboration~\cite{Abe:2003jk}.

A new lattice study of $D_s$  mesons
by the UKQCD Collaboration has appeared recently~\cite{Dougall:2003hv} and
a paper by Terasaki~\cite{Terasaki:2003qa}
on the new $D_s$ mesons was submitted to the preprint server
only one day after this article.

In view of the possibility of similar states in the $B_s$ spectrum
it appears
worthwhile to mention that the static $n_f=2$ lattice results presented
here imply that the scalar quark model $B_s$ meson should have a
mass of 5837(43)(24) MeV, with additional $1/m$ corrections of
order 40~MeV, possibly upwards, based on the $D_s$ experience.
This has to be compared with the $B\overline{K}$ threshold of about 5775 MeV.

\begin{acknowledgments}
This work has been supported by the EU network
HPRN-CT-2000-00145, a PPARC Advanced Fellowship
(grant PPA/A/S/2000/00271) and PPARC grants PPA/G/O/2000/00454,
PPA/G/O/2002/00463.
GSB thanks Ismail Zahed for discussions.
He also thanks the SESAM/T$\chi$L Collaboration, in particular
Thorsten Struckmann, Norbert Eicker, Boris Orth,
Bram Bolder and Thomas Lippert for their contribution and
acknowleges useful comments from Ted Barnes, Frank Close, Thorsten Feldmann,
Chris Maynard and
Sheldon Stone.
\end{acknowledgments}


\begin{thebibliography}{}
\bibitem{Aubert:2003fg}
B.~Aubert {\it et al.}  [BaBar Collaboration],
%``Observation of a narrow meson decaying to D/s+ pi0 at a mass of  2.32-GeV/c**2,''
Phys.\ Rev.\ Lett.\  {\bf 90}, 242001 (2003)
[arXiv:hep-ex/0304021].
%%CITATION = HEP-EX 0304021;%%

\bibitem{Besson:2003jp}
D.~Besson  [CLEO Collaboration],
%``Observation of a Narrow Resonance of Mass 2.46 GeV/c^2 in the D_s^*+\pi^0 Final State, and Confirmation of the D_sJ^*(2317)^*,''
arXiv:hep-ex/0305017.
%%CITATION = HEP-EX 0305017;%%

\bibitem{Cahn:2003cw}
R.~N.~Cahn and J.~D.~Jackson,
%``Spin-orbit and tensor forces in heavy-quark light-quark mesons:  Implications of the new D/s state at 2.32-GeV,''
arXiv:hep-ph/0305012.
%%CITATION = HEP-PH 0305012;%%

\bibitem{Barnes:2003dj}
T.~Barnes, F.~E.~Close and H.~J.~Lipkin,
%``Implications of a D K molecule at 2.32-GeV,''
arXiv:hep-ph/0305025.
%%CITATION = HEP-PH 0305025;%%

\bibitem{Godfrey:wj}
S.~Godfrey and R.~Kokoski,
%``The Properties Of P Wave Mesons With One Heavy Quark,''
Phys.\ Rev.\ D {\bf 43}, 1679 (1991);
%%CITATION = PHRVA,D43,1679;%%
S.~Godfrey and N.~Isgur,
%``Mesons In A Relativized Quark Model With Chromodynamics,''
Phys.\ Rev.\ D {\bf 32} (1985) 189.
%%CITATION = PHRVA,D32,189;%%

\bibitem{Cheng:2003kg}
H.~Y.~Cheng and W.~S.~Hou,
%``B decays as spectroscope for charmed four-quark states,''
Phys.\ Lett.\ B {\bf 566}, 193 (2003)
[arXiv:hep-ph/0305038].
%%CITATION = HEP-PH 0305038;%%

\bibitem{Szczepaniak:2003vy}
A.~P.~Szczepaniak,
%``Description of the D/s*(2320) resonance as the D pi atom,''
arXiv:hep-ph/0305060.
%%CITATION = HEP-PH 0305060;%%

\bibitem{vanBeveren:2003kd}
E.~van Beveren and G.~Rupp,
%``Observed D/s(2317) and tentative D(2030) as the charmed cousins of the  light scalar nonet,''
Phys.\ Rev.\ Lett.\  {\bf 91}, 012003 (2003)
[arXiv:hep-ph/0305035].
%%CITATION = HEP-PH 0305035;%%

\bibitem{Bardeen:2003kt}
W.~A.~Bardeen, E.~J.~Eichten and C.~T.~Hill,
%``Chiral multiplets of heavy-light mesons,''
arXiv:hep-ph/0305049.
%%CITATION = HEP-PH 0305049;%%

\bibitem{nowak}
M.~A.~Nowak, M.~Rho and I.~Zahed,
%``Chiral effective action with heavy quark symmetry,''
Phys.\ Rev.\ D {\bf 48}, 4370 (1993)
[arXiv:hep-ph/9209272];
%%CITATION = HEP-PH 9209272;%%
D.~Ebert, T.~Feldmann, R.~Friedrich and H.~Reinhardt,
%``Effective meson Lagrangian with chiral and heavy quark symmetries from quark flavor dynamics,''
Nucl.\ Phys.\ B {\bf 434}, 619 (1995)
[arXiv:hep-ph/9406220];
%%CITATION = HEP-PH 9406220;%%
D.~Ebert, T.~Feldmann and H.~Reinhardt,
%``Extended NJL model for light and heavy mesons without q anti-q  thresholds,''
Phys.\ Lett.\ B {\bf 388}, 154 (1996)
[arXiv:hep-ph/9608223].
%%CITATION = HEP-PH 9608223;

\bibitem{Bardeen:1993ae}
W.~A.~Bardeen and C.~T.~Hill,
%``Chiral dynamics and heavy quark symmetry in a solvable toy field theoretic model,''
Phys.\ Rev.\ D {\bf 49}, 409 (1994)
[arXiv:hep-ph/9304265].
%%CITATION = HEP-PH 9304265;%%

\bibitem{Colangelo:2003vg}
P.~Colangelo and F.~De Fazio,
%``Understanding $D_{sJ}(2317)$,''
arXiv:hep-ph/0305140.
%%CITATION = HEP-PH 0305140;%%

\bibitem{Godfrey:2003kg}
S.~Godfrey,
%``Using Radiative Transitions to Test the 1^3P_0(c\bar{s}) Nature of the D_{sJ}^*(2317)^+ State,''
arXiv:hep-ph/0305122.
%%CITATION = HEP-PH 0305122;%%

\bibitem{Michael:1998sg}
C.~Michael and J.~Peisa  [UKQCD Collaboration],
%``Maximal variance reduction for stochastic propagators with applications  to the static quark spectrum,''
Phys.\ Rev.\ D {\bf 58}, 034506 (1998)
[arXiv:hep-lat/9802015].
%%CITATION = HEP-LAT 9802015;%%

\bibitem{Alexandrou:1992ti}
C.~Alexandrou, S.~G\"usken, F.~Jegerlehner, K.~Schilling and R.~Sommer,
%``The Static approximation of heavy - light quark systems: A Systematic lattice study,''
Nucl.\ Phys.\ B {\bf 414}, 815 (1994)
[arXiv:hep-lat/9211042];
%%CITATION = HEP-LAT 9211042;%%

\bibitem{Duncan:1994uq}
A.~Duncan, E.~Eichten, J.~Flynn, B.~Hill, G.~Hockney and H.~Thacker,
%``Properties of B mesons in lattice QCD,''
Phys.\ Rev.\ D {\bf 51}, 5101 (1995)
[arXiv:hep-lat/9407025].
%%CITATION = HEP-LAT 9407025;%%

\bibitem{Ewing:1995ih}
A.~K.~Ewing {\it et al.}  [UKQCD Collaboration],
%``Heavy quark spectroscopy and matrix elements: a lattice study using the static approximation,''
Phys.\ Rev.\ D {\bf 54}, 3526 (1996)
[arXiv:hep-lat/9508030].
%%CITATION = HEP-LAT 9508030;%%

\bibitem{Bali:1998pi}
G.~S.~Bali and P.~Boyle,
%``A lattice potential investigation of quark mass and volume dependence  of the Upsilon spectrum,''
Phys.\ Rev.\ D {\bf 59}, 114504 (1999)
[arXiv:hep-lat/9809180];
%%CITATION = HEP-LAT 9809180;%%
G.~S.~Bali,
%``QCD forces and heavy quark bound states,''
Phys.\ Rept.\  {\bf 343}, 1 (2001)
[arXiv:hep-ph/0001312].
%%CITATION = HEP-PH 0001312;%%

\bibitem{Bali:2000vr}
G.~S.~Bali {\it et al.}  [SESAM Collaboration],
%``Static potentials and glueball masses from QCD simulations with Wilson  sea quarks,''
Phys.\ Rev.\ D {\bf 62}, 054503 (2000)
[arXiv:hep-lat/0003012];
%%CITATION = HEP-LAT 0003012;%%
B.~Bolder {\it et al.} [SESAM Collaboration],
%``A high precision study of the Q anti-Q potential from Wilson loops in  the regime of string breaking,''
Phys.\ Rev.\ D {\bf 63} (2001) 074504
[arXiv:hep-lat/0005018].
%%CITATION = HEP-LAT 0005018;%%

\bibitem{Hein:2000qu}
J.~Hein {\it et al.},
%``Scaling of the B and D meson spectrum in lattice QCD,''
Phys.\ Rev.\ D {\bf 62}, 074503 (2000)
[arXiv:hep-ph/0003130].
%%CITATION = HEP-PH 0003130;%%

\bibitem{Lewis:2000sv}
R.~Lewis and R.~M.~Woloshyn,
%``S and P-wave heavy-light mesons in lattice NRQCD,''
Phys.\ Rev.\ D {\bf 62}, 114507 (2000)
[arXiv:hep-lat/0003011].
%%CITATION = HEP-LAT 0003011;%%

\bibitem{Boyle:1997rk}
P.~Boyle  [UKQCD Collaboration],
%``Heavy meson spectroscopy at beta = 6.0,''
Nucl.\ Phys.\ Proc.\ Suppl.\  {\bf 63}, 314 (1998)
[arXiv:hep-lat/9710036];
%%CITATION = HEP-LAT 9710036;%%
P.~Boyle  [UKQCD Collaboration],
%``Comprehensive Spectroscopy Of D, D/S And Charmonium Systems From  Relativistic Lattice QCD,''
Nucl.\ Phys.\ Proc.\ Suppl.\  {\bf 53} (1997) 398.
%%CITATION = NUPHZ,53,398;%%
\bibitem{Michael:1999nq}
C.~Michael and P.~Pennanen  [UKQCD Collaboration],
%``Two heavy-light mesons on a lattice,''
Phys.\ Rev.\ D {\bf 60}, 054012 (1999)
[arXiv:hep-lat/9901007].
%%CITATION = HEP-LAT 9901007;%%
\bibitem{Abe:2003jk}
K.~Abe {\it et al.} [Belle Collaboration],
%``Measurements of the D/sJ resonance properties,''
arXiv:hep-ex/0307052 and
%%CITATION = HEP-EX 0307052;%%
%``Observation of the D/sJ*(2317) and D/sJ*(2460) in B decays,''
arXiv:hep-ex/0307041.
%%CITATION = HEP-EX 0307041;%%
\bibitem{Dougall:2003hv}
A.~Dougall, R.~D.~Kenway, C.~M.~Maynard and C.~McNeile  [UKQCD
                  Collaboration],
%``The spectrum of D/s mesons from lattice QCD,''
arXiv:hep-lat/0307001.
%%CITATION = HEP-LAT 0307001;%%
\bibitem{Terasaki:2003qa}
K.~Terasaki,
%``BABAR resonance as a new window of hadron physics,''
Phys.\ Rev.\ D {\bf 68}, 011501 (2003)
[arXiv:hep-ph/0305213].
%%CITATION = HEP-PH 0305213;%%


\end{thebibliography}
\end{document}